\documentclass[conference]{IEEEtran}
\IEEEoverridecommandlockouts

\usepackage{cite}
\usepackage{amsmath,amssymb,amsfonts}
\usepackage{algorithmic}
\usepackage{graphicx}
\usepackage{subcaption}
\usepackage{hyperref}
\usepackage{textcomp}
\usepackage{xcolor}
\usepackage{comment}
\def\BibTeX{{\rm B\kern-.05em{\sc i\kern-.025em b}\kern-.08em
    T\kern-.1667em\lower.7ex\hbox{E}\kern-.125emX}}
\begin{document}

\title{Towards Robust Semantic Video Transmission over Block Erasure Channels
}

\author{\IEEEauthorblockN{Nargis Fayaz\textsuperscript{1}, Homa Esfahanizadeh\textsuperscript{2}, Matin Mortaheb\textsuperscript{2}, Jinfeng Du\textsuperscript{2}, Harish Viswanathan\textsuperscript{2}}
\IEEEauthorblockA{\textsuperscript{1}\textit{Nokia Bell Labs, India} \quad
\textsuperscript{2}\textit{Nokia Bell Labs, Murray Hill, New Jersey, USA} \\
eez218533@ee.iitd.ac.in, \{homa.esfahanizadeh, matin.mortaheb, jinfeng.du, harish.viswanathan\}@nokia-bell-labs.com}
}

\maketitle

\begin{abstract}
This paper investigates semantic-aware neural joint source-channel coding (JSCC) for robust video transmission over block erasure channels. We propose a neural video compression framework exploring both spatial-domain and feature-domain designs. In the spatial domain, video frames are partitioned into blocks, enabling localized erasure handling and fine-grained robustness control via uniform erasure and two-level, semantic-guided non-uniform erasure strategies. In the feature domain, latent features are partitioned, enabling missing features to be semantically recovered while maintaining overall spatial consistency. Comprehensive experiments quantify reconstruction quality under varying uniform and non-uniform erasure probabilities. Our results show that spatial-domain JSCC excels at handling random localized losses, whereas feature-domain JSCC provides superior robustness to distributed erasures and maintains fidelity under low-loss scenarios. The analysis highlights the trade-offs between spatial continuity and semantic redundancy, offering insights for designing robust, task-aware video communication systems.
\end{abstract}

\begin{IEEEkeywords}
Block erasure channels, joint source-channel coding, robust video transmission, semantic communication, video compression.\vspace{-0.1cm}
\end{IEEEkeywords}

\section{Introduction}
Video dominates global internet traffic, accounting for over $80\%$ of downstream data, driven by applications such as video-on-demand, conferencing, and immersive media. The increasing adoption of ultra-high-definition formats (4K/8K), high frame rates, and high dynamic range substantially raises bitrate requirements, making efficient compression essential for scalable storage and transmission. Traditional codecs such as H.264/AVC~\cite{wiegand2003overview} and H.265/HEVC~\cite{sullivan2012overview} employ hybrid block-based frameworks that integrate intra and inter prediction, transform coding, quantization, and entropy coding. End-to-end neural network based video compression, using autoencoders, variational autoencoders, or recurrent and transformer architectures, jointly optimizes spatiotemporal feature extraction, quantization, and entropy coding, often surpassing HEVC in rate–distortion performance~\cite{joy2023deep,gomes2025end,ding2021advances}.

Beyond compression, reliable video transmission over practical wireless channels requires robustness to noise, fading, and packet loss. Classical systems address compression and error correction separately, following the Shannon separation principle, which can be inefficient in lossy environments and practical code lengths. Neural {joint source channel coding} (JSCC) overcomes this limitation by directly mapping source video to channel symbols and reconstructing it at the receiver, enabling graceful quality degradation over a noisy channel ~\cite{tung2022deepwive,bourtsoulatze19}. This paradigm is particularly advantageous for real-time applications such as video conferencing, telemedicine, and UAV streaming, where retransmission is often impractical or ineffective due to low-delay requirements.

Recently, semantic video communication has emerged as a complementary paradigm.  Applications include video conferencing~\cite{jiang2022wireless}, autonomous driving~\cite{wang2023semantic}, and general video understanding using rate-distortion autoencoders~\cite{Habibian_2019_ICCV}. Integrating semantic awareness with NN-based JSCC enables a unified framework that jointly optimizes compression, semantic relevance, and channel robustness, making it well suited for low-latency, resource-constrained, and lossy communication environments~\cite{wang2022wireless}.

Building on the growing interest in semantic-aware and end-to-end neural JSCC for video transmission, we focus on enhancing reliability over heterogeneous block erasure channels. Related work has investigated image transmission over binary symmetric channels \cite{Split_JSCC_TCOM}.
The remainder of this paper is organized as follows. Section~II presents the system model and the proposed JSCC-based video transmission framework under block erasure channels. Section~III analyzes spatial-domain block partitioning, including uniform and semantic-guided non-uniform erasure strategies, and evaluates their impact on robustness and reconstruction quality. Section~IV studies feature-domain block partitioning and examines its resilience to uniform and non-uniform erasures through experimental analysis. Finally, Section~V concludes the paper.

\section{System Model and Method}
We consider a reliable transmission of compressed video over a multi-level block erasure channel. 
During transmission, a random subset of the constituent blocks of a video representation may be erased due to channel impairments or bandwidth constraints. The decoder must reconstruct the frame from a subset of successfully received blocks.

\subsection{Semantic-Aware Neural Joint Source--Channel Coding}

To mitigate the impact of block erasures, we build a semantic-aware JSCC mechanism upon the end-to-end deep video compression (DVC) framework proposed in~\cite{lu2019dvc}. The DVC architecture comprises motion estimation, motion compensation, and residual coding modules that jointly exploit temporal and spatial redundancies. An entropy estimation module learns the probability model of the latent representations to estimate the coding rate (in bits per pixel, BPP), which is combined with the distortion term to drive the rate–distortion optimization.

In our design, the JSCC encoder maps the latent representation into $K$ structured blocks, introducing controlled redundancy to improve robustness against erasures.
Let $\{B_i\}_{i=1}^K$ denote the $K$ blocks generated by the encoder. To enable the decoder to recover from block erasures, during the code design, each block $B_i$ is transmitted over an independent erasure channel $C_{\epsilon_i}(\cdot)$ with the erasure probability $\epsilon_i$, defined as
\begin{equation}
C_{\epsilon_i}(B_i) =
\begin{cases}
B_i, & \text{with probability } 1 - \epsilon_i, \\[4pt]
[-1]_{\mathrm{dim}(B_i)}, & \text{with probability } \epsilon_i.
\end{cases}
\end{equation}
If a block is erased, a fixed placeholder filled with $-1$ replaces the erased block at the decoder's input. Since latent features lie within $[0,1]$, the value $-1$ enables reliable identification of erased blocks while preserving a fixed input dimension for the decoder. This will inherently make our solution a JSCC for the multi-level block erasure channel.

Two partitioning strategies are considered for generating blocks: (i) \textit{spatial-domain block partitioning} (Section~\ref{SD}), where motion and residual tensors are segmented into localized spatial regions, and (ii) \textit{feature-domain block partitioning} (Section~\ref{sec:feature_erasure}), where partitioning is performed along the channel dimension to distribute semantic information across feature groups.
To reflect their relative contribution to reconstruction quality, each block is assigned an importance score that quantifies the expected distortion increase in the event of its loss.
We train the code over a predefined multi-level block erasure channel. During training, erasure of the $i$-th block, $i\in\{1,\dots,K\}$, is implemented by sampling an independent binary random variable $r_i$ with
\(
\Pr(r_i = 0) = \epsilon_i
\)
and computing
\(
C_{\epsilon_i}(B_i) = r_i (B_i + 1) - 1.
\)
This formulation enables differentiable simulation of erasures while maintaining compatibility with gradient-based optimization.

\begin{figure}[htbp]
\centering
\includegraphics[width=1\linewidth]{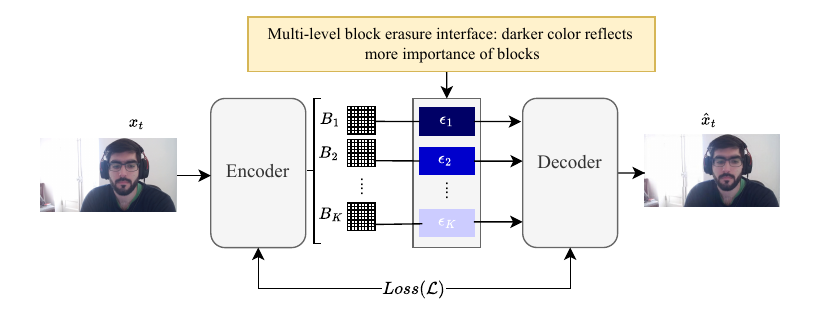}
\caption{End-to-end JSCC-based video transmission system.}
\label{fig: system model}
\end{figure}

During training, an entropy model estimates the bit rate of the quantized latents to guide the rate-distortion optimization. The motion and hyperprior latents are modeled by learned distributions, while the residual latent is coded under a prior whose parameters are predicted from the hyperprior. The resulting differentiable rate estimate, measured in BPP, is combined with the reconstruction distortion and minimized end-to-end, where a trade-off parameter balances rate against quality. At inference, the latents are uniformly quantized and entropy coded into the transmitted bitstream. The decoder then reconstructs the frame from the subset of successfully received blocks, exploiting the learned redundancy and contextual dependencies to mitigate the effect of missing information (see Fig.~\ref{fig: system model}).

We consider two erasure mechanisms: (i) \textit{uniform erasure}, where $\epsilon_i = \epsilon$ for all $i$, modeling random packet loss or bandwidth constraints, and (ii) \textit{non-uniform erasure}, where erasure probabilities are assigned according to block importance, reflecting congestion-aware packet prioritization. Training under these settings equips the model with resilience to random and structured channel impairments.

The network is trained end-to-end on the Vimeo-90k dataset~\cite{xue2019video} using the Adam optimizer with an initial learning rate of $10^{-4}$, batch size $8$, and input resolution $256 \times 256$, for $50$ epochs. The rate-distortion trade-off is controlled by $\lambda = 2048$. A sigmoid activation constrains the latents to the $[0,1]$ range, after which they are quantized to 
$8$-bit precision before entropy coding. 
The learning rate is reduced by a factor of $0.8$ when the validation loss plateaus (patience of $3$ epochs). Evaluation is performed on the Video Conferencing Dataset (VCD)~\cite{naderi2024vcd}, which contains 160 talking-head videos across four scenarios: raw webcam (TH), opaque background (TH-OB), blurred background (TH-BB), and handheld mobile (THM). Each scenario includes 40 sequences ($\geq 30\%$ active speakers) in YUV420p, 1080p, 30~FPS, 10s format.

\section{Spatial Domain Block Partitioning}
\label{SD}

For spatial domain block partitioning, the motion and residual tensors produced by the encoder are divided into $16$ blocks of size $ C \times H/4 \times W/4$, where $C$ is the number of channels, and $H$ and $W$ represent the frame height and width, respectively. In this setting, erasures correspond to missing localized regions. The number of channels is not the same for motion and residual tensors.

Recovery under this regime leverages the inherent properties of convolutional neural networks, which exploit translation invariance and spatial correlations among neighboring regions. When trained with spatial erasures, the network learns to in-paint missing regions by aggregating contextual cues from surrounding un-erased blocks, aided by redundancy across feature channels that encode overlapping low-level structures.

\subsection{Uniform Spatial Domain Erasure}
\label{sec:spatial_performance}
\begin{figure}
\centering
\includegraphics[width=0.9\linewidth]{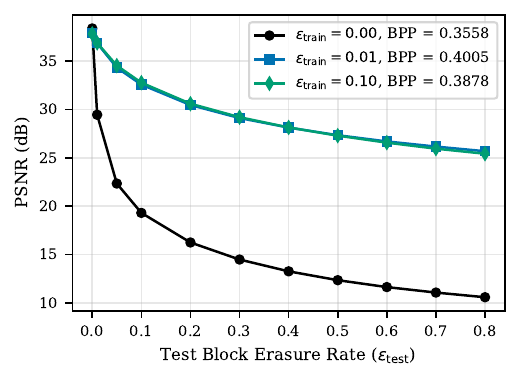}
\caption{Mismatch analysis: PSNR for models trained with different $\epsilon_{\text{train}}$ and tested under varying $\epsilon_{\text{test}}$ (Spatial domain).}
\label{fig:vcd_mismatch_bler_cleanTX_spatial_uniform}
\end{figure}

Here, we evaluate the performance of the proposed system under uniform spatial-domain erasure, where partitions are defined in the pixel domain of each frame, and each partition has the same erasure probability. We first perform a mismatch analysis in Fig.~\ref{fig:vcd_mismatch_bler_cleanTX_spatial_uniform}, evaluating codes trained under $\epsilon_{\text{train}}$ and tested at $\epsilon_{\text{test}}$. The erasure‑ignorant 
baseline achieves the highest PSNR on a clean channel but degrades sharply as erasures appear. In contrast, the erasure‑aware models sacrifice only a fraction of a dB at $\epsilon_{\text{test}} = 0$ yet stay above $25$~dB even when $80\%$ of blocks are erased. Notably, their performance is largely insensitive to $\epsilon_{\text{train}}$ because spatial partitioning reduces recovery to an in-painting problem where the decoder reconstructs each missing block from its spatial neighbors and temporal context, a local skill the network acquires from only a small amount of training erasure. This contrasts with channel partitioning (next section), where increasing $\epsilon_{\text{train}}$ does improve robustness. Overall, exposing the model to erasures during training trades a negligible clean-channel penalty for a large gain in packet-loss resilience. We note the curves are not rate-matched, with each model settling at a slightly different bitrate.

\begin{figure}
\centering
\setlength{\tabcolsep}{0pt}
\renewcommand{\arraystretch}{0.4}
\newlength{\rowlab}\setlength{\rowlab}{1.6em}
\newlength{\panelw}\setlength{\panelw}{\dimexpr(\columnwidth-\rowlab)/3\relax}
\newlength{\panelh}\setlength{\panelh}{\dimexpr\panelw*1024/1920\relax}
\newcommand{\rowlabel}[1]{%
  \makebox[\rowlab][c]{%
    \raisebox{0.5\panelh}[\panelh][0pt]{\rotatebox[origin=c]{90}{\scriptsize #1}}}}
\begin{tabular}{@{}c@{}c@{}c@{}c@{}}
 & {\scriptsize Lossy frame} & {\scriptsize Frame $+1$} & {\scriptsize Frame $+2$} \\[1pt]
\rowlabel{Base}
  & \includegraphics[width=\panelw]{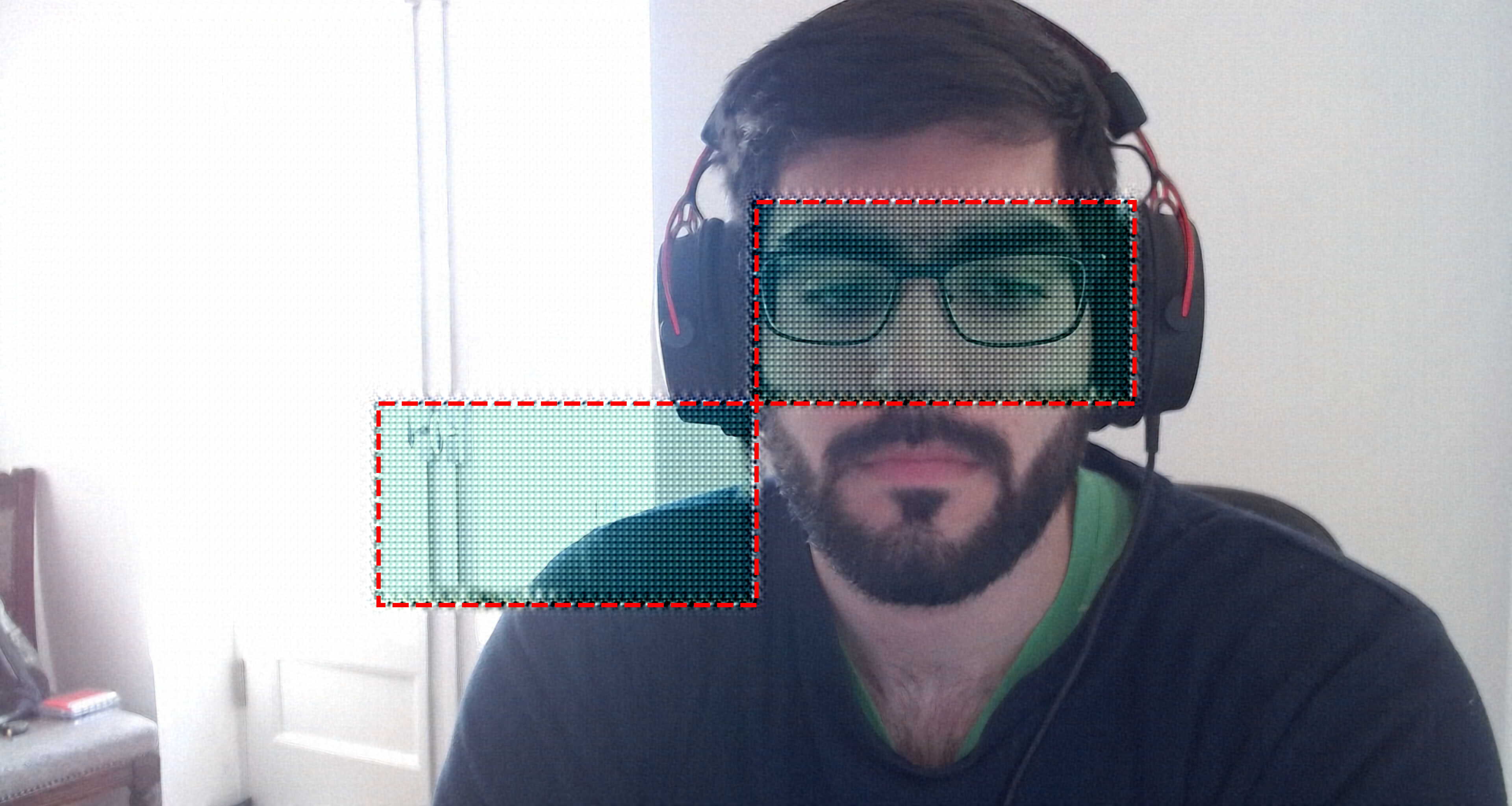}
  & \includegraphics[width=\panelw]{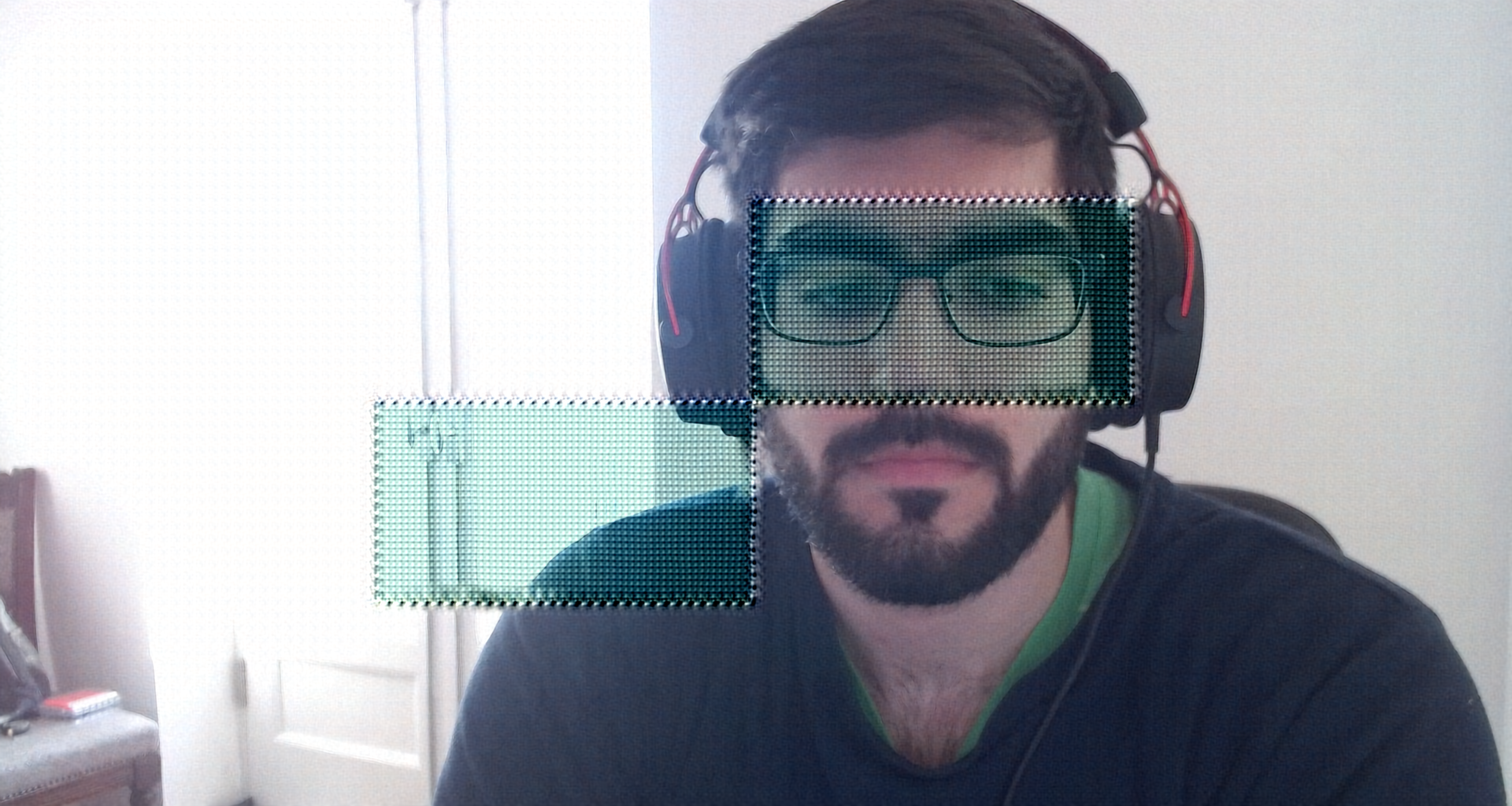}
  & \includegraphics[width=\panelw]{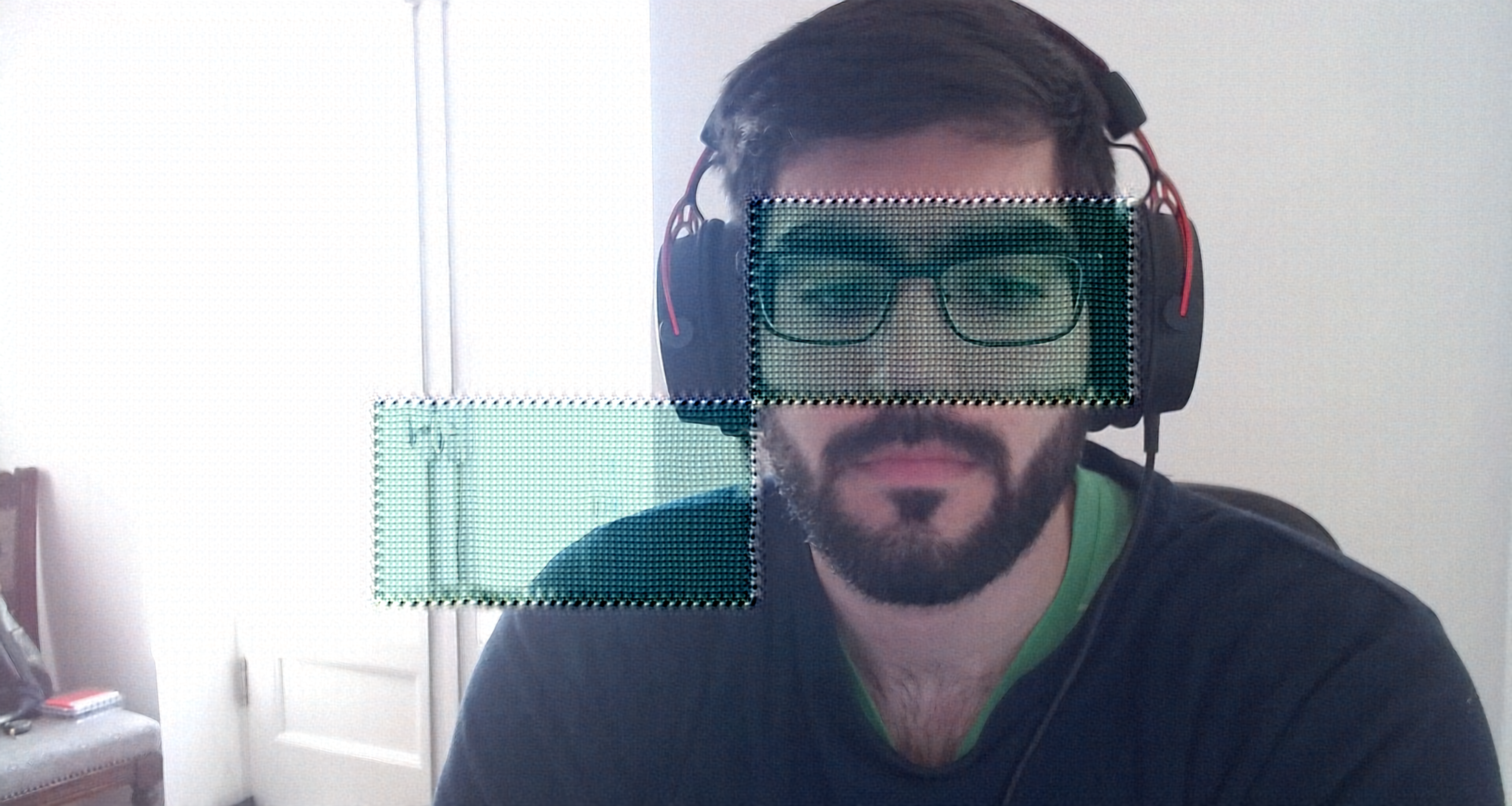} \\
\rowlabel{Unif. $0.05$}
  & \includegraphics[width=\panelw]{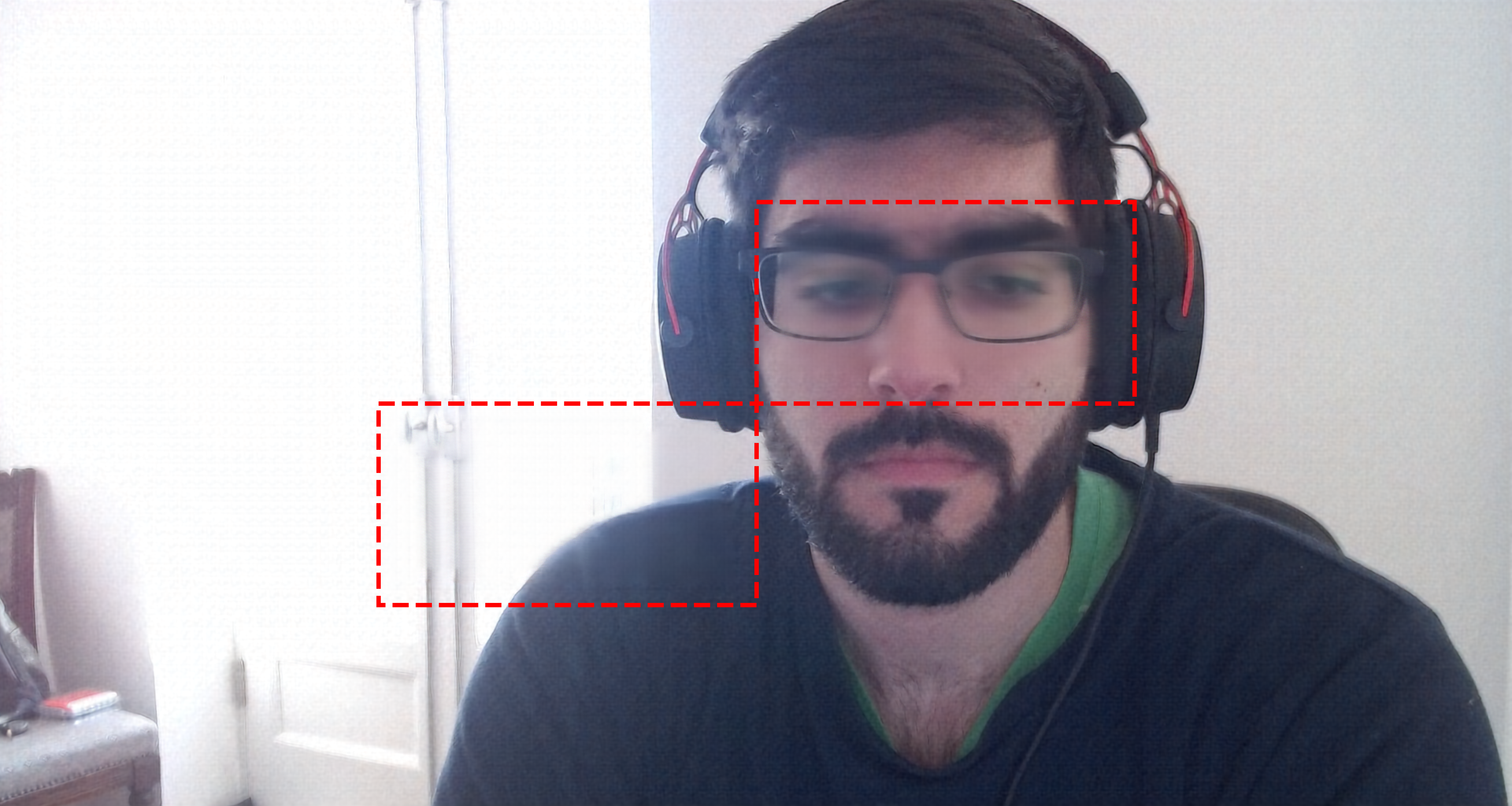}
  & \includegraphics[width=\panelw]{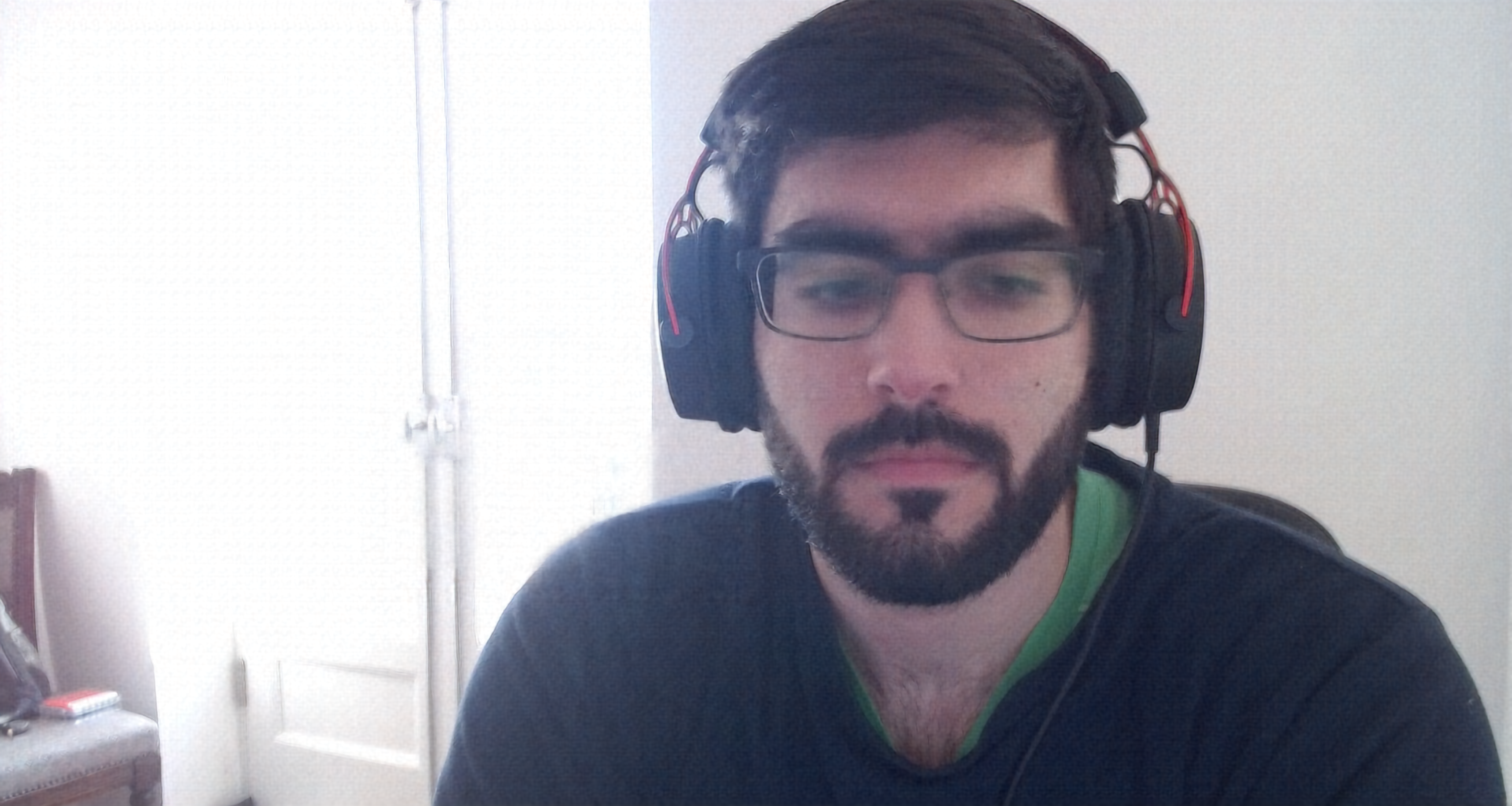}
  & \includegraphics[width=\panelw]{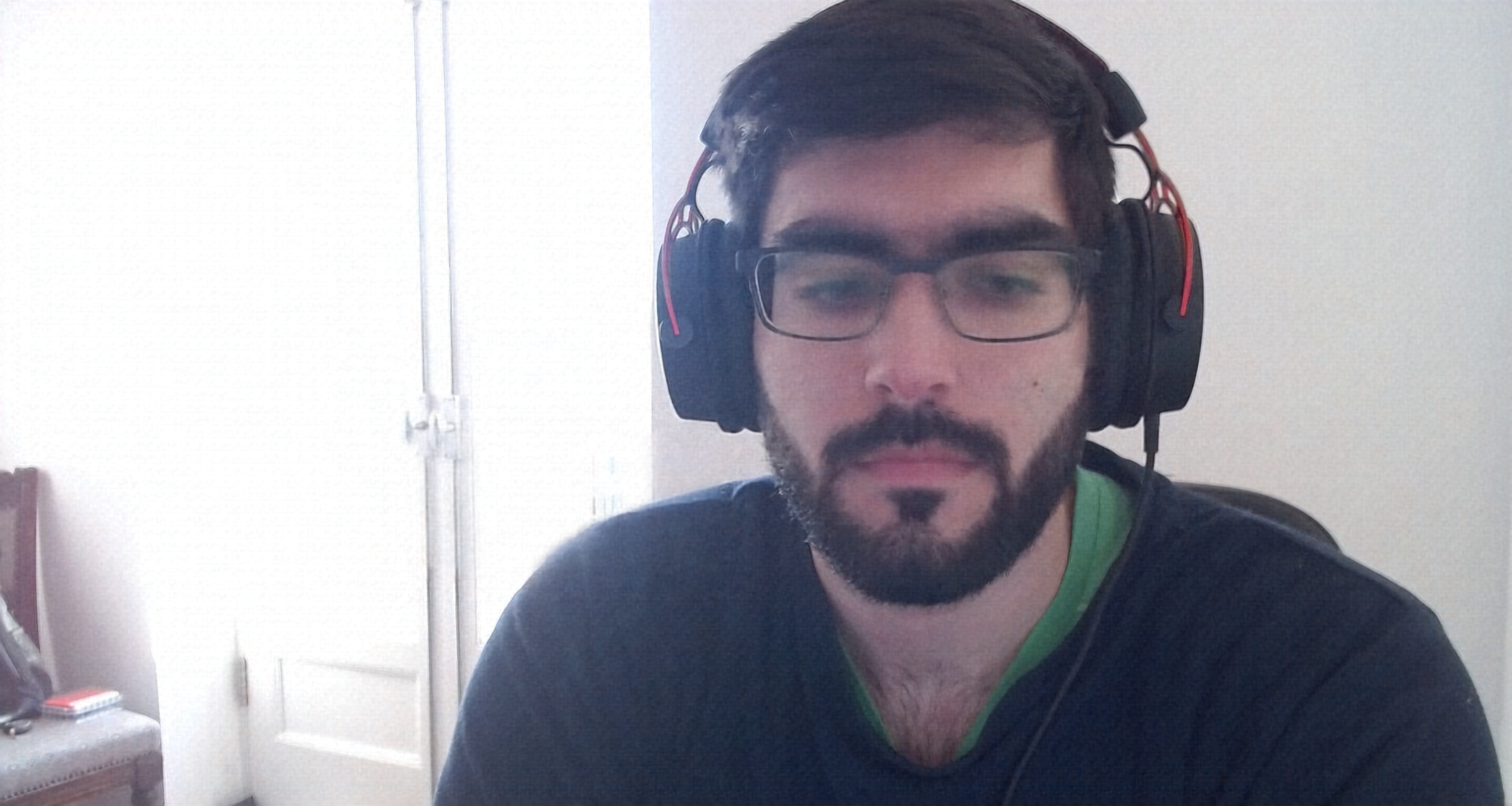} \\
\rowlabel{Unif. $0.10$}
  & \includegraphics[width=\panelw]{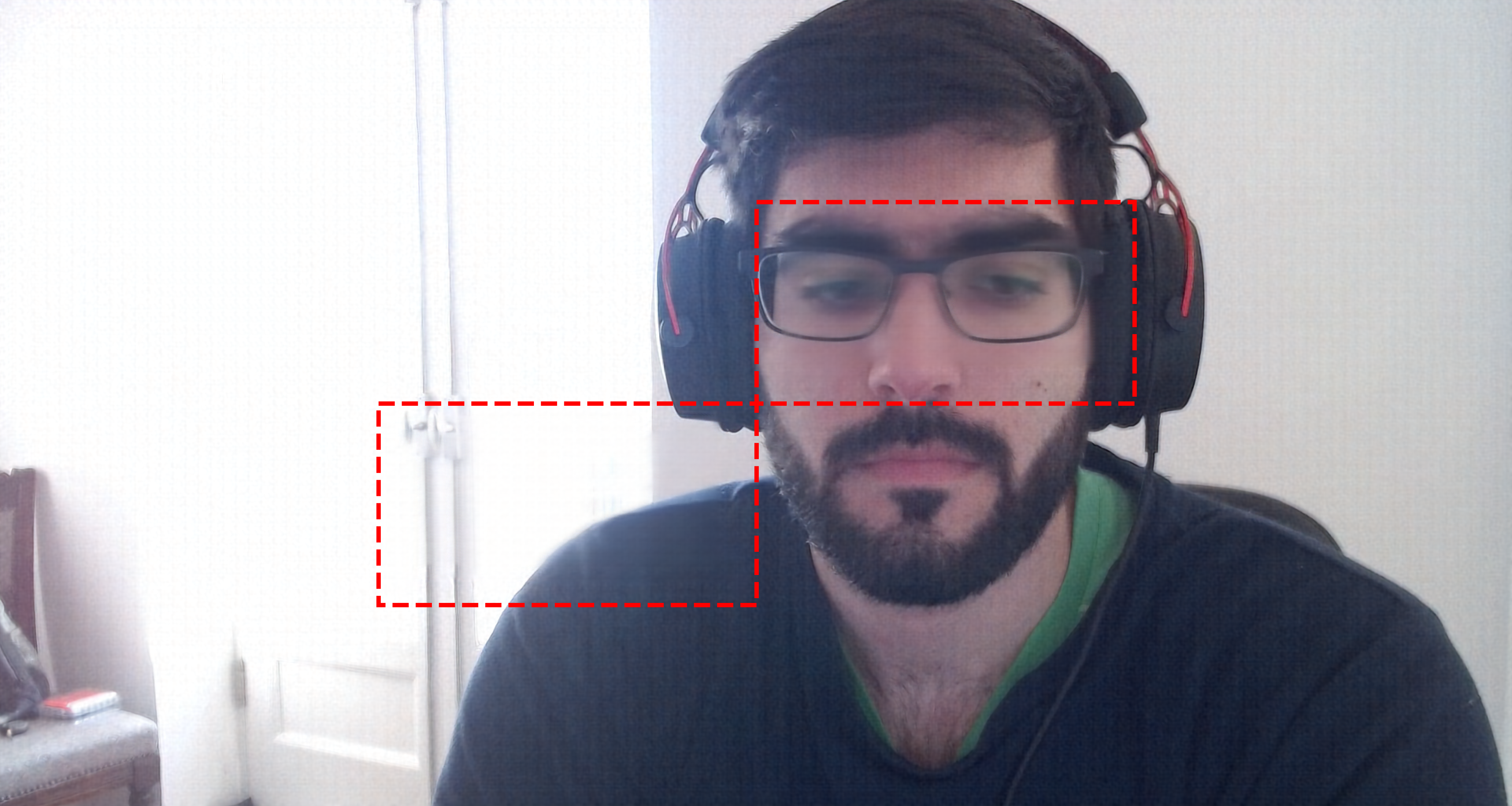}
  & \includegraphics[width=\panelw]{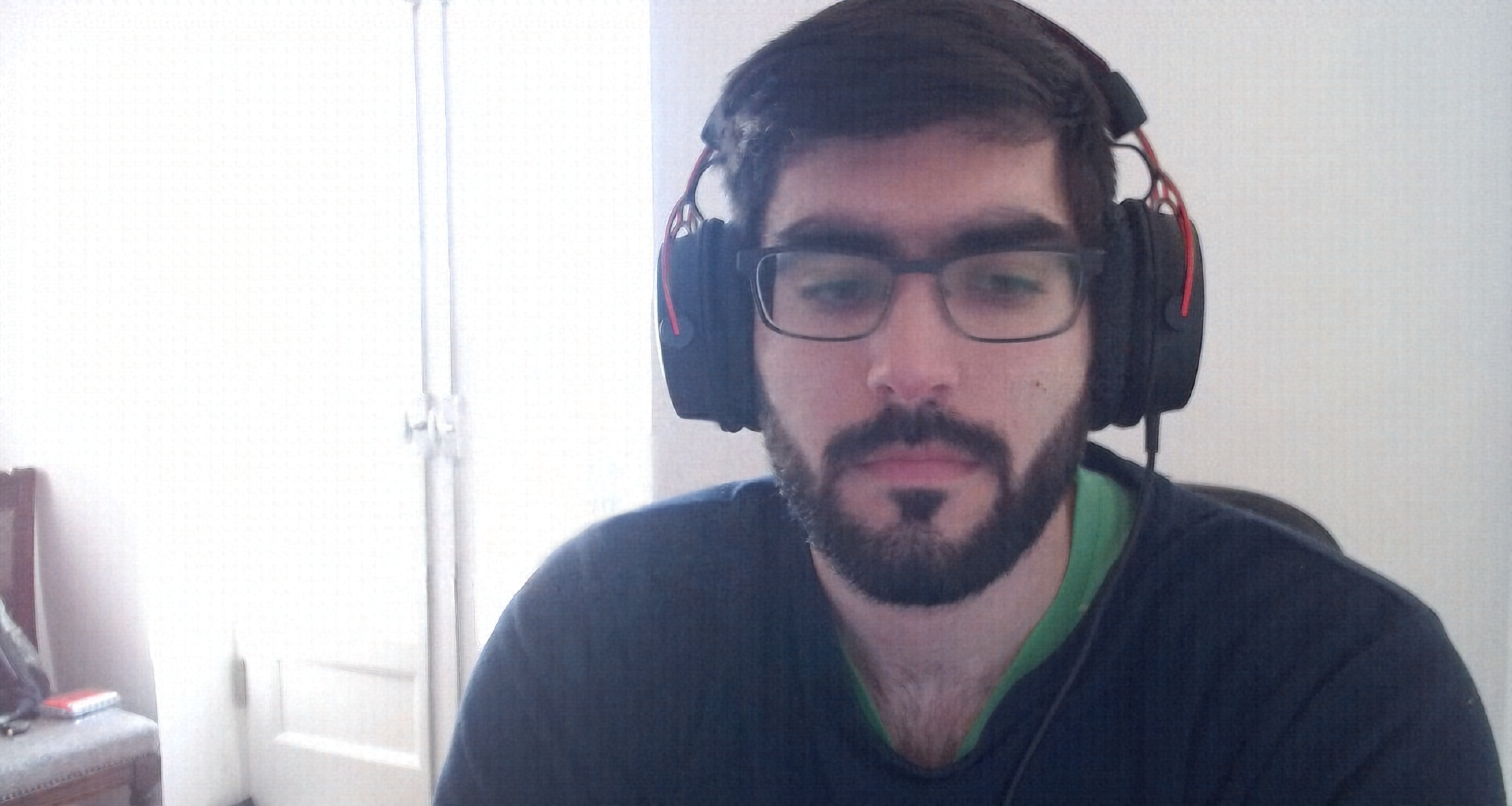}
  & \includegraphics[width=\panelw]{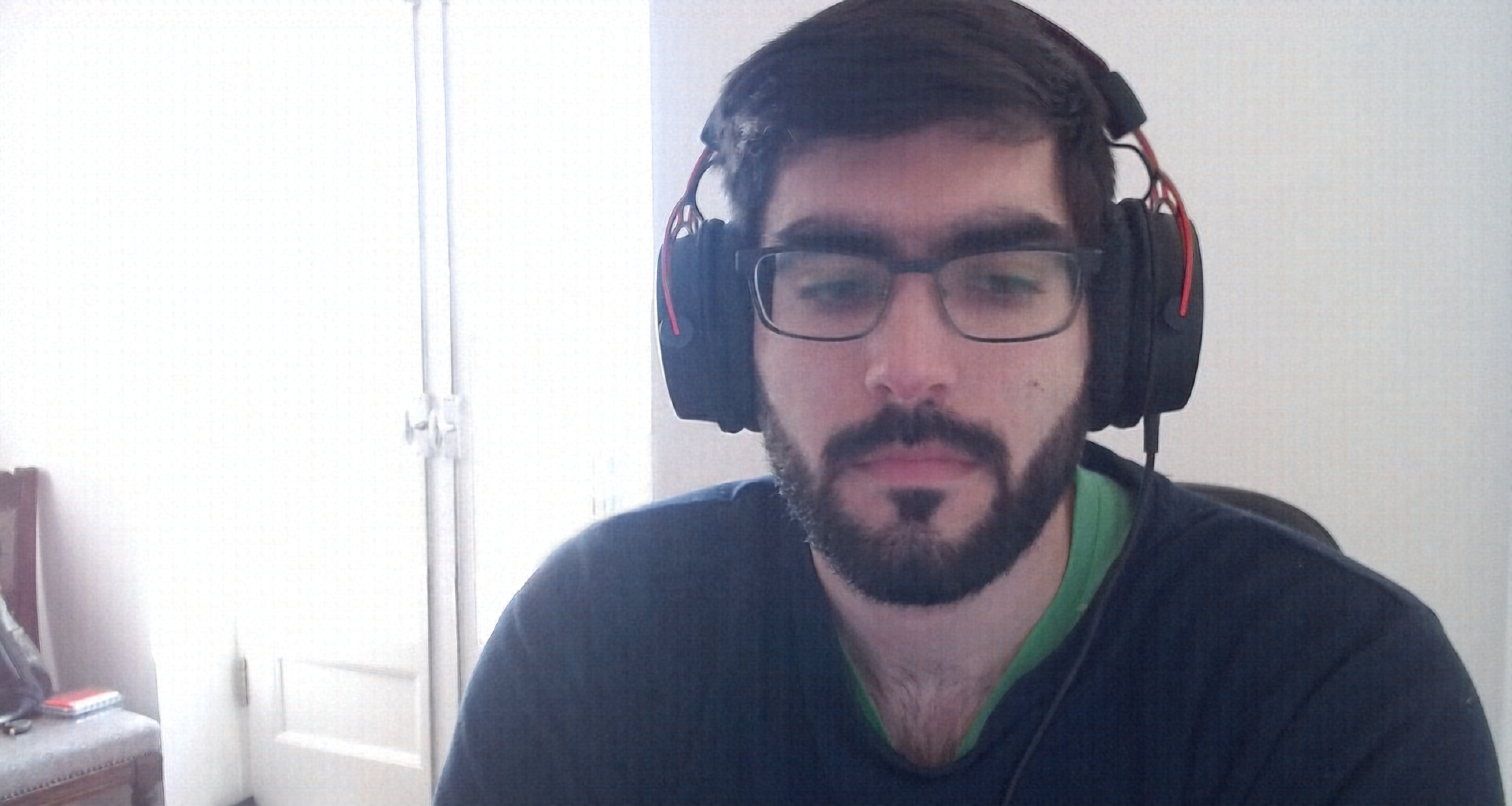} \\
\end{tabular}
\caption{{Visual comparison between the base and erasure-aware uniform JSCC systems under a transient erasure of the $7^{\text{th}}$ and $10^{\text{th}}$ blocks on the third frame
(outlined in red), with the two subsequent recovered frames.}}
\label{fig:visual_comparison_spatial}
\end{figure}

Fig.~\ref{fig:visual_comparison_spatial} presents a visual comparison between the base system and the erasure-aware JSCC system (trained with uniform block erasure) under a transient erasure of the $7^{\text{th}}$ and $10^{\text{th}}$ blocks on the third frame, followed by the subsequent recovery frames (no other erasure occurs elsewhere). The base system fails to reconstruct the erased blocks, and the resulting degradation persists as the corrupted frame is used as the reference for subsequent frames, whereas the erasure-aware system effectively exploits spatial redundancy to recover the lost blocks and rapidly restores quality over the following frames. Quantitatively, averaged over the three frames, the full-frame PSNR improves from $22.99$~dB for the base system to $37.91$~dB and $38.54$~dB for the erasure-aware system trained at erasure probabilities of $0.05$ and $0.10$, respectively.

\subsection{Two-level Spatial Domain Erasure}
\label{sec:2spatial_performance}

A two-level spatial erasure framework is proposed, guided by spatial importance within the video frame. In video conferencing scenarios, the central regions that typically contain the face and upper body have higher semantic relevance compared to the peripheral areas. Thus, each frame is divided into $16$ spatial blocks, with lower erasure probability ($0.05$) assigned to the semantically important central blocks and higher erasure probability ($0.1$) to the peripheral ones. This task-aware design provides greater protection to critical content while allowing higher losses in background regions, thereby enhancing robustness under spatially structured erasures.

\begin{figure}
\centering
\includegraphics[width=0.9\linewidth]{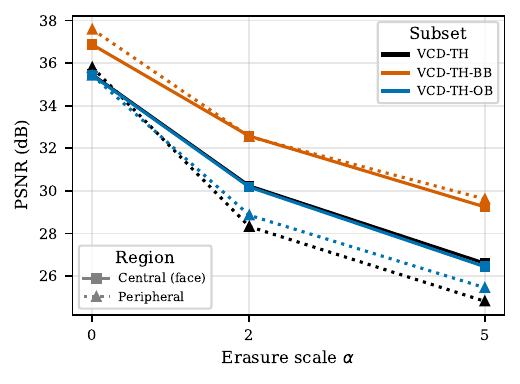}
\caption{Regional PSNR under varying levels of non-uniform erasure across different VCD categories (BPP for VCD-TH: $0.4188$, VCD-TH-BB: $0.3765$, VCD-TH-OB: $0.4229$).}
\label{fig:VCD_categories}
\end{figure}

The performance of the proposed system under non-uniform erasures across different categories of the VCD dataset is shown in Fig.~\ref{fig:VCD_categories}, where we report the reconstruction PSNR separately for the central (face) and peripheral regions. The test erasures are applied as multiples of the training erasures in the two-level scheme, with $\epsilon_{\text{test}} = \alpha \cdot \epsilon_{\text{train}}$, where $\alpha \in \{0, 2, 5\}$. Under clean conditions ($\alpha = 0$), the two regions perform comparably, with the central face region marginally lower since it carries more high-frequency detail and motion than the background. As the erasure scale increases, the benefit of the importance-aware allocation emerges: for VCD-TH and VCD-TH-OB, whose backgrounds contain genuine detail that erasure degrades, the protected central region pulls ahead of the periphery by roughly $1$--$2$~dB at $\alpha = 5$. In contrast, VCD-TH-BB attains the highest overall PSNR and shows almost no regional gap across all $\alpha$. This is due to the fact that its blurred and low-resolution background is both easy to code and inherently resilient to erasure, so the periphery never becomes the weak point and the protection mechanism has little to compensate for.

\section{Feature Domain Block Partitioning}
\label{sec:feature_erasure}
For feature-domain block partitioning, the motion and residual tensors generated by the encoder are divided into eight blocks along the channel dimension, each of size $C/8 \times H \times W$. In this feature-space erasure framework, erasures correspond to the loss of entire feature groups, while the full spatial resolution is preserved. Each block typically encodes a subset of semantic attributes, such as texture, boundaries, or object-level cues.

Recovery in this setting relies on the inherent redundancy across channels, as different channel groups are designed to have overlapping semantic information. By training with random block erasures, the network learns cross-channel mappings that approximate missing feature groups from the preserved ones, effectively performing semantic imputation rather than spatial inpainting. Importantly, since the correlation across spatial field remains intact, global consistency is maintained in the reconstruction, enabling the decoder to produce semantically coherent outputs.

Unlike spatial-domain systems, the feature-domain erasure rate during training makes a pronounced difference in erasure resilience at test time. This is because the task is no longer reduced to spatial in-painting (i.e., reconstructing missing pixels from neighboring spatial context). Instead, the network learns to exploit cross-channel correlations, envisioning the content of lost feature groups from the semantic redundancy retained in the surviving channels. As a result, models exposed to feature-domain erasure during training are more robust than their spatial-domain counterparts when the test erasure rate is small. More quantitative results characterizing this trade-off are discussed in the subsequent subsections.

\subsection{Performance Evaluation over Uniform and Non-Uniform Block Erasure Channels}

\begin{figure}
    \centering
    \includegraphics[width=0.9\linewidth]{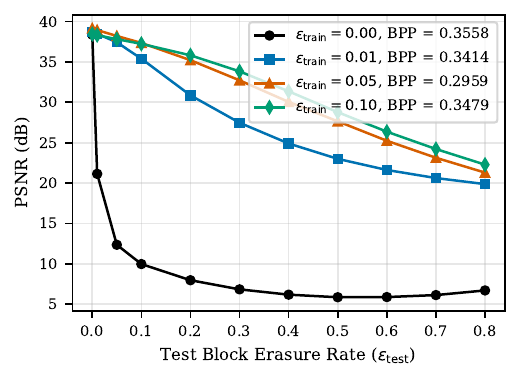}
    \caption{Mismatch analysis: PSNR for models trained with different $\epsilon_{\text{train}}$ and tested under varying $\epsilon_{\text{test}}$ (Feature domain).}
    \label{fig:video_mismatch}
\end{figure}

We perform a feature-domain mismatch analysis similar to that in the spatial domain (see Fig.~\ref{fig:video_mismatch}). 
The baseline (un-trained for erasures) suffers a severe performance loss even at very low $\epsilon_{\text{test}}$, dropping from $38.37$~dB in the clean case to $21.14$~dB at just $\epsilon_{\text{test}} = 0.01$ and below $8$~dB by $\epsilon_{\text{test}} = 0.2$, indicating its inability to recover lost features. 
Models trained with feature erasures exhibit substantially improved robustness: $\epsilon_{\text{train}} = 0.01$ provides moderate resilience with a gradual decline (from $38.48$~dB at $\epsilon_{\text{test}} = 0.01$ to $23.00$~dB at $\epsilon_{\text{test}} = 0.5$) and $\epsilon_{\text{train}} = 0.10$ achieves the highest robustness, showing minimal degradation up to $\epsilon_{\text{test}} \approx 0.3$ (retaining $33.83$~dB) with a smooth decline beyond. Notably, this robustness comes at negligible clean-channel cost, as all three models remain within a fraction of a dB of the baseline's clean PSNR at $\epsilon_{\text{test}} = 0$. We note, however, that the models settle at slightly different operating rates (all lower than the baseline) during training, so the curves are not perfectly rate-matched, and one robust model ($\epsilon_{\text{train}} = 0.05$) in fact operates at the lowest BPP, confirming that the resilience gains stem from the erasure-aware training rather than from any additional bit budget.

\begin{figure}
    \centering
    \includegraphics[width=0.9\linewidth]{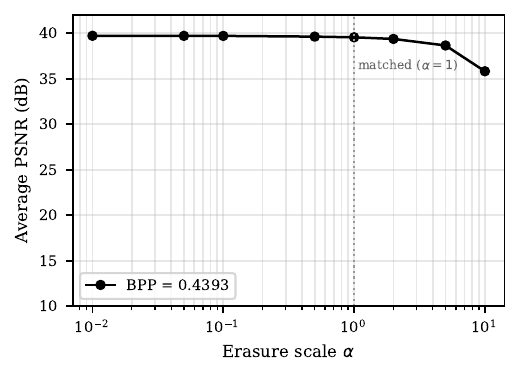}
   \caption{Video performance under non-uniform erasure, trained with $\epsilon_{\text{train}} = [0, 0.01, 0.02, 0.03, 0.04, 0.05, 0.06, 0.07]$ and tested with $\epsilon_{\text{test}}=\alpha \epsilon_{\text{train}}$ and varying $\alpha$.}
    \label{fig:video_non_uniform}
\end{figure}

For the non-uniform erasure scenario, both the motion and residual blocks of the video compression system were trained with progressively increasing erasure probabilities, $\epsilon_{\text{train}} = [0, 0.01, 0.02, 0.03, 0.04, 0.05, 0.06, 0.07]$. The trained models were evaluated across multiple test erasure levels, defined as $\epsilon_{\text{test}} = \alpha \cdot \epsilon_{\text{train}}$, where $\alpha \in \{0.01, 0.05, 0.1, 0.5, 1, 2, 5, 10\}$. The results (see Fig.~\ref{fig:video_non_uniform}) indicate that the system maintains high reconstruction quality across a wide range of erasure levels, with the average PSNR staying near its peak of about $39.7$~dB for all test erasures at or below the training range ($\alpha \le 1$). At the matched operating point ($\alpha = 1$) the PSNR is $39.5$~dB, and it decreases only gradually as the test erasure grows beyond the training range. Even when the test erasure exceeds the training range by up to ten times ($\alpha = 10$), the average PSNR remains above $35$~dB, demonstrating strong robustness and effective recovery capability. This graceful degradation highlights the system's ability to generalize to previously unseen high erasure conditions. These observations confirm that progressive erasure training effectively ensures both high-fidelity reconstruction and resilience under extreme non-uniform erasures.

\begin{figure}[htbp]
\centering
\includegraphics[width=0.8\linewidth]{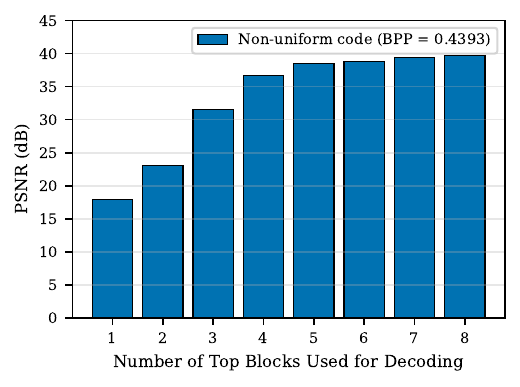}
\caption{{Non-uniform training over feature-domain block erasures ranks blocks by importance, enabling promising reconstruction from only the top-$k$ most important blocks.}}
\label{fig:nu_decoding}
\end{figure}

Fig.~\ref{fig:nu_decoding} presents the average PSNR as a function of the number of decoding blocks in the video reconstruction process. PSNR increases monotonically with more decoding blocks, rising sharply from approximately $18$~dB with a single block to around $37$~dB by the fourth block, and gradually approaching $40$~dB when all eight blocks are used. The system is trained under non-uniform erasure conditions, encouraging unequal error protection (UEP). Early decoding blocks carry critical coarse-level information and are better protected, while later blocks encode finer residuals and refinements. The steep gains in PSNR in the first few blocks, followed by diminishing improvements, reflect the hierarchical importance of information across blocks. Under bandwidth constraints, only half of the decoding blocks may be transmitted for reconstruction. Even in this limited scenario, UEP ensures that essential information from early blocks is preserved, enabling high-quality reconstruction despite partial block usage.

\begin{figure}[htbp]
\centering
\includegraphics[width=0.99\linewidth]{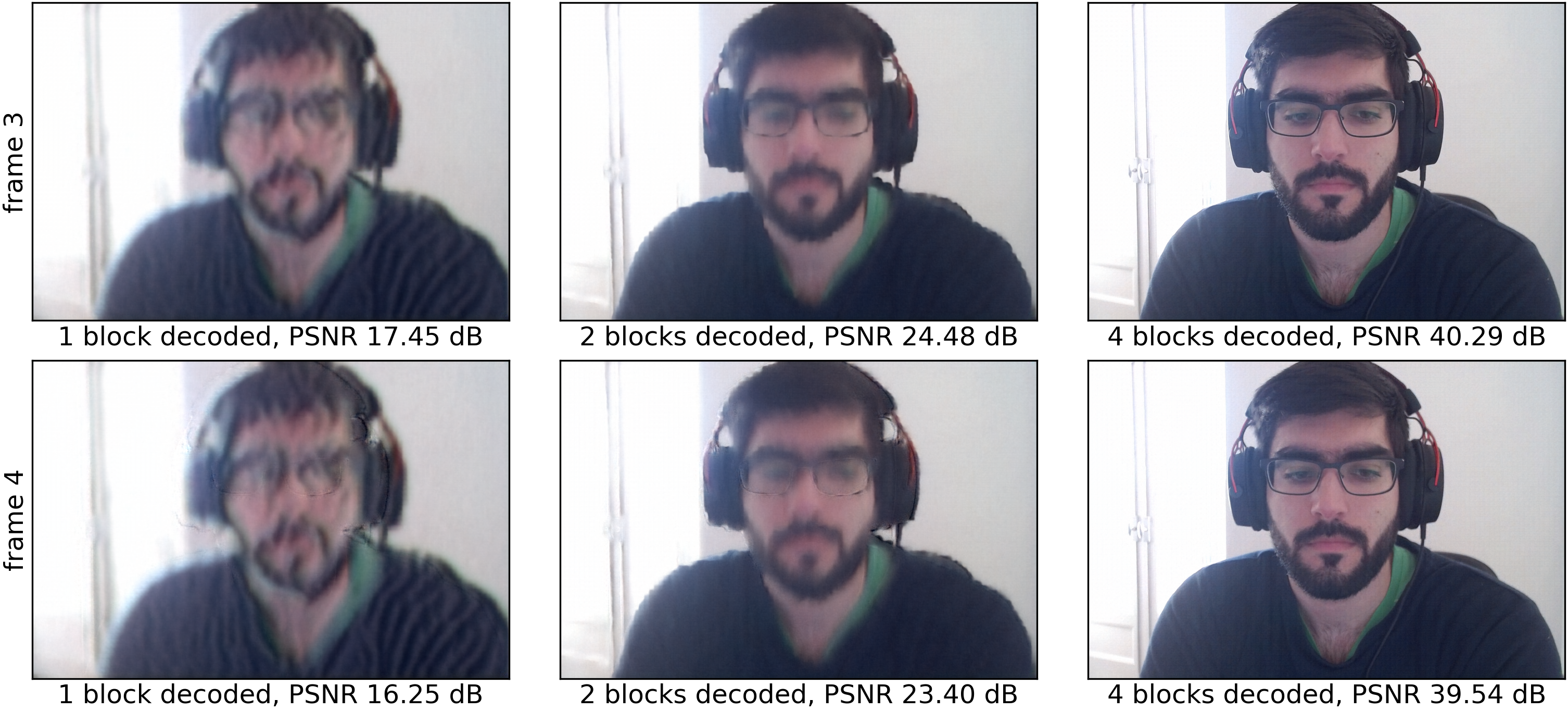}
\caption{{Reconstruction quality under non-uniform feature-domain erasure when only the most important blocks are received. Blocks are ordered by importance, so even one block preserves essential information and more blocks refine detail. 
}}
\label{vcd_blocks_comparison}
\end{figure}

Fig.~\ref{vcd_blocks_comparison} provides a visual illustration of this behavior, highlighting the system's ability to maintain high-quality outputs despite partial block reception and severe erasure conditions.  Errors propagate along the group of frames, seen as the small PSNR drop from the third to the fourth frame.

\section{Conclusion}
This work explored semantic-aware neural JSCC for robust video transmission over block erasure channels. We analyzed two complementary strategies: spatial-domain partitioning, which enables fine-grained control and semantic-guided protection, and feature-domain partitioning, which leverages semantic redundancy for enhanced resilience to distributed losses. 
The proposed framework aligns well with the requirements of next-generation low-latency networks, such as 6G and beyond. Its ability to adaptively allocate protection without explicit semantic exchange or feedback enables efficient support for real-time applications, including video conferencing, teleoperation, and robotic control in unreliable channel conditions.

\bibliographystyle{IEEEtran}
\bibliography{ref}

\end{document}